\documentstyle[prl,aps]{revtex}
\twocolumn
\begin{document}
\input{epsf}
\draft
\twocolumn[\hsize\textwidth\columnwidth\hsize\csname@twocolumnfalse\endcsname
\title{Dynamics of fractal dimension during phase
ordering of a geometrical multifractal}
\author{Avner Peleg \cite{adr1} and Baruch Meerson}
\address{The Racah Institute  of  Physics, Hebrew
University   of  Jerusalem,
Jerusalem 91904, Israel}
\maketitle
\begin{abstract}
A simple multifractal coarsening model is suggested that can
explain the observed dynamical behavior of the fractal
dimension in a wide range of coarsening fractal systems.
It is assumed that the minority
phase (an ensemble of droplets) at $t=0$ represents a
non-uniform recursive fractal set, and
that this set is a geometrical multifractal
characterized
by a $f(\alpha)$-curve. It is assumed that
the droplets shrink according to their size
and preserving their ordering.
It is shown that at early times the Hausdorff
dimension does
not change with time, whereas at late times its dynamics follow the
$f(\alpha)$ curve.
This
is illustrated by a special
case of a two-scale Cantor
dust. The results are then generalized to
a wider range of coarsening mechanisms.
\end{abstract}
\pacs{PACS numbers: 64.60.Ak, 61.43.Hv}
 \vskip1pc]
\narrowtext
Fractal growth phenomena have been under extensive investigation
during the past two decades \cite{Feder,Vicsek,Meakin2}.
The inverse process of fractal coarsening
occurs in many physical systems. It has been discussed
in the context of sintering of fractal matter \cite{Jullien1}.
Coarsening of fractal clusters by surface tension
in the bulk-diffusion-controlled \cite{Irisawa1,Conti}, interface-controlled
\cite{Irisawa2} and edge-diffusion-controlled \cite{Irisawa1,Jullien2}
systems has been investigated. Additional examples
include thermal relaxation of rough
grain boundaries \cite{Streitenberger} and smoothing
of fractal polymer structure in the process of polymer collapse
\cite{Crooks}. Two-dimensional
fractal fingering, observed in a Hele-Shaw cell with
radial symmetry (for a review see Ref. \onlinecite{fingering}),
exhibits coarsening at a late
stage of the experiment. All these systems are quite different, as they
involve non-conserved or conserved
order parameters, different transport mechanisms, etc.

A crucial issue related to any phase ordering process is the presence
or absence of dynamical
scale invariance (DSI) \cite{Gunton}. DSI assumes that
there is a single dynamical
length scale $\lambda (t)$ such that the coarsening system
looks (statistically) invariant in time when lengths are scaled by
$\lambda (t)$. Does a fractal cluster or a fractal interface
exhibit DSI (on a shrinking
interval of distances) in the process of
coarsening? Early scenarios of fractal coarsening in systems with
non-conserved \cite{Toyoki} and conserved \cite{Jullien1} order parameter
did rely upon the hypothesis of DSI.
However, numerical simulations showed that the
DSI breaks down during the coarsening of fractal
clusters
in edge- \cite{Jullien2}
and bulk-diffusion-controlled \cite{Conti} systems. On the
other hand, recent simulations of
smoothing of a fractal
polymer during collapse \cite{Crooks},
and of interface-controlled fractal
coarsening under a global conservation law \cite{Peleg1},
do support DSI. Therefore, a question
arises about possible universality classes of fractal
coarsening.

Even if DSI holds, the fractal dimension
may or may not change with time.
Early fractal coarsening
scenarios \cite{Toyoki,Jullien1} assumed that it
remains constant (again,
on a shrinking interval of distances).
Experiments on
sintering of silica aerogels (a convenient way of investigating
fractal coarsening) have been inconclusive. Some
of them \cite{Jullien1}
gave evidence in favor of constancy of the fractal dimension
during coarsening, while others \cite{Hinic} reported a
significant change of
the fractal dimension with time.
Another evidence for a significant decrease of fractal dimension
with time was found in experiments on thermal annealing
of ferroelectric thin films of lead zirconate titanate \cite{Shur}.
In this experiment, the fractal dimension remained constant at early
times, and decreased to its final value at intermediate times.
Numerical simulations of a variety of coarsening systems
with different growth laws
showed that the fractal dimension does not change with time.
These simulations include bulk-diffusion-controlled
\cite{Irisawa1,Conti}, edge-diffusion-controlled
\cite{Irisawa1,Jullien2}, and interface-controlled
\cite{Irisawa2,Peleg1} systems.

It is remarkable that in so many systems with widely different
coarsening mechanisms the fractal dimension remains constant during
the dynamics. Therefore, one is tempted to look for a general scenario
that would explain this fact and that would be insensitive to
specific coarsening mechanisms. The simple multifractal coarsening model
developed in this paper has this property. In addition, this
model is the first attempt to address the multifractal properties
of fractal coarsening.

We shall consider a
very simple model of a coarsening fractal system.
In this model, the initial condition for the minority
phase is an ensemble of
droplets that represents a {\it geometrical multifractal}.
We will then assume that
the smaller
droplets shrink and disappear independently, according to their sizes,
and consider discrete time dynamics.
Using a well-known theorem of multifractal geometry,
we will establish the
dynamical behavior of the Hausdorff
dimension of this simple coarsening system.
This result will be illustrated in
a special case, when the droplets are distributed in the
form of a two-scale Cantor dust \cite{Feder,Vicsek}.
Employing the size distribution function of this
fractal set\cite{Peleg},
we will follow the dynamical behavior of the $d$-measure in two
characteristic
limiting cases and show that
the Hausdorff dimension's dynamics in this example
are consistent with
the general result. Then we will relax the discrete time assumption.
Furthermore, we will show that the results are
essentially independent of the details of the
coarsening dynamics as long as the minority-phase droplets do
not merge or break up.

The minority phase of our
model represents, at zero time, a big but finite
ensemble of droplets that form
a non-uniform recursive fractal \cite{Vicsek} with a constant density
distribution in the $E$-dimensional space.
Let us index the droplets in the $m$-th generation of the fractal
according to their radii. Thus, all the droplets with index $k$
have radius $R_{m}(k)$ and form a subset of the whole fractal
which we denote by $S_{m}(k)$. The smallest droplets have index
$k=0$ and radius $R_{m}(0)$, which is the lower cutoff of the
fractal. The largest droplets have index $k=m$ and radius
$R_{m}(m)$, which is the upper cutoff. One
can work with a size distribution function $n_{m}(k)$,
which is simply the number of droplets with radius $R_{m}(k)$,
and use it to compute the Hausdorff dimension of
the fractal (see Ref. \cite {Peleg},
where this was done for
a two-scale Cantor dust).

Any non-uniform recursive fractal with a constant
density distribution can be described as multifractal in
geometrical sense (see Ref. \cite {Vicsek}, p. 66).
In this case one can introduce the measure
of the subset $S_{m}(k)$ in the following way:
\begin{equation}
\mu_{m}(k)=\frac{R_{m}^{E}(k)}{\Sigma_{k=0}^{m} n_{m}(k)R_{m}^{E}(k)}
\,,
\label{1}
\end{equation}
where $R_{m}(k)$ are the radii of
the droplets divided by the size of the system.
The H\"{o}lder
exponent of the elements of the subset $S_{m}(k)$ is defined
by \cite {Halsey}
\begin{equation}
\alpha_{m}(k)=\frac{\ln \mu_{m}(k)}{\ln R_{m}(k)}
\,.
\label{2}
\end{equation}
The $f(\alpha)$
curve for the fractal is constructed in the following way
\cite {Feder,Vicsek,Mandelbrot2}:
\begin{equation}
f(\alpha)=-\frac{\ln n_{m}(k)}{\ln R_{m}(k)} \;\;\;\;\;\;
(1 \ll k \ll m)
\,,
\label{3}
\end{equation}
where $k$ is supposed to be expressed through $\alpha$
with the help of the equation $\alpha_{m}(k)=\alpha$. (We assume 
that this equation gives a one-to-one
correspondence between $\alpha$ and $k$.)
$f(\alpha)$ is assumed to have
a single maximum which is attained for $\alpha=\alpha_{0}$,
so that $f(\alpha_{0})$ is the Hausdorff dimension
of the whole fractal. We also assume that $f(\alpha(k))$
is the Hausdorff dimension of the subset $S_{m}(k)$.
This assumption, widely used in the physical literature,
was rigorously proved in the case of a two-scale Cantor dust
\cite {Billingsley}, and also for a class of other multifractal
measures \cite {Peyriere}.

We now turn to describe the dynamics. We assume first that the
droplets shrink and disappear independently, according to their radius
only, and also simplify the governing
dynamics by introducing a discrete time $\tau$ (later we will relax these
two assumptions).
In the first time step $\tau=0$ the
smallest droplets with radius $R_{m}(0)$ disappear,
{\it while the sizes of the other droplets do not change}.
In the next time step $\tau=1$ the elements with radius
$R_{m}(1)$ disappear, and so on.
The set of droplets that survive after each step of these
dynamics obviously remains
self-similar (on a shrinking interval of distances).
The main result of this paper is
the following behavior of the
Hausdorff dimension $D$ as a function of the discrete time
$\tau$. For $\tau \leq k(\alpha_{0})$ $D$ does not
change: $D(\tau)=D_{0}$, where $D_{0}$ is
the Hausdorff dimension of the initial condition.
For $\tau >  k(\alpha_{0})$ $D(\tau)=f(\alpha(k_{min}))$
where $k_{min} (\tau)$ is the $k$-value of the smallest
droplets which have not yet disappeared by
time $\tau$.
This dynamical behavior is illustrated in Fig. 1.
\begin{figure}[h]
\vspace{0.0cm}
\hspace{-0.5cm}
\rightline{ \epsfxsize = 8.0cm \epsffile{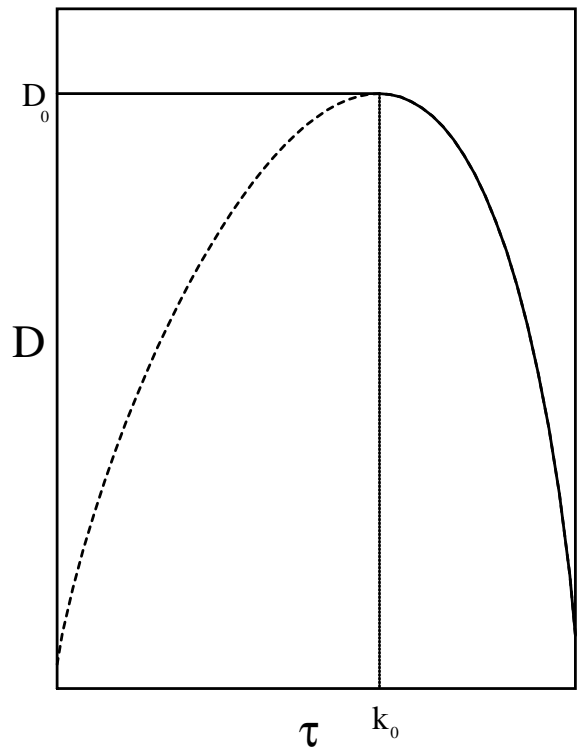}}
\caption{
Hausdorff dimension $D$ of the ensemble of droplets
versus discrete time $\tau$ (solid curve). The dashed
curve is the $f(\alpha (k))$ curve at $\tau=0$.
$D_{0}$ is the Hausdorff dimension at $\tau=0$,
while $k_{0} \equiv k(\alpha_{0})$ is the value of $k$ for
which $f(\alpha (k))$ has its maximum.
\label{Fig. 1}}
\end{figure}

The proof of this result is based on the following
theorem: {\it the Hausdorff dimension of a
union of two fractal sets $S_{1}$ and $S_{2}$ with fractal
dimensions $D_{S_{1}}$ and $D_{S_{2}}$, respectively,
is $D=\mbox {max}\,(D_{S_{1}},D_{S_{2}})$}. (See, for example,
\cite {Vicsek}, p. 17.)
In the last time step of the dynamics, $\tau=m$, the coarsening
object consists of the subset $S_{m}(m)$ alone,
and its Hausdorff dimension is $f(\alpha(m))$.
In the previous time step $\tau=m-1$ the object consists
of two subsets : $S_{m}(m)$ with Hausdorff dimension
$f(\alpha(m))$, and $S_{m}(m-1)$ with Hausdorff dimension
$f(\alpha(m-1))$. It follows from the shape of the $f(\alpha)$
curve of the initial fractal that
$f(\alpha(m-1))>f(\alpha(m))$. Using the
theorem, we get $D(\tau=m-1)=f(\alpha (m-1))$.

More generally, consider time $\tau=k_{0}+s$,
where $s$ is a positive integer and
$k_{0}\equiv k(\alpha_{0})$. At this time we can regard the
object as a union of two fractal subsets
$S_{m}(k_{0}+s)$ and $S_{m}(m \geq k\geq k_{0}+s+1)$. Here,
$S_{m}(m \geq k\geq k_{0}+s+1)$ is the union of all
subsets $S_{m}(k)$ with $k=s+1,...,m$. It is also the whole
coarsening object at time $\tau=k_{0}+s+1$.
Assume by induction that
$D(\tau=k_{0}+s+1)=D(S_{m}(m \geq k\geq k_{0}+s+1))=f(\alpha(k_{0}+s+1))$.
It follows from the shape of the $f(\alpha)$ curve that
$f(\alpha(k_{0}+s)) > f(\alpha(k_{0}+s+1))$.
Hence, using the theorem, we
conclude that $D(\tau=k_{0}+s)=f(\alpha(k_{0}+s))$.
Since $k_{0}+s$ is the index of the smallest droplets
which have not yet disappeared we can write this result as
\begin{equation}
D(\tau \geq k_{0}) = f(\alpha(k_{min}))
\,.
\label{8}
\end{equation}

The dynamical behavior of the  Hausdorff dimension at times
$\tau \leq k_{0}$ can be found in a similar way. For $\tau=k_{0}-1$
the object can be considered as a union of
two fractal subsets $S_{m}(m\geq k\geq k_{0})$ and
$S_{m}(k_{0}-1)$. It follows from Eq. (\ref {8}) that
$D(\tau=k_{0})=D(S_{m}(m\geq k\geq k_{0}))=D_{0}$. From the
shape of the $f(\alpha)$ curve we get $D_{0}=f(\alpha(k_{0})) >
f(\alpha(k_{0}-1))$. Therefore, $D(\tau=k_{0}-1)=D_{0}$. More generally,
for any time $\tau=k_{0}-s$ the coarsening object can be considered
as a union of the two fractal subsets :
$S_{m}(m\geq k\geq k_{0}-s+1)$ with Hausdorff dimension $D_{0}$
and $S_{m}(k_{0}-s)$ with Hausdorff dimension $f(\alpha(k_{0}-s))$. 
From the shape of the $f(\alpha)$ curve we deduce
$f(\alpha(k_{0}-s))<D_{0}$.
Hence, by using the above theorem
we conclude that $D(\tau=k_{0}-s)=D_{0}$.
More generally, we can write:
\begin{equation}
D(\tau \leq k_{0}) = f(\alpha(k_{0}))=D_{0}
\,.
\label{11}
\end{equation}

Let us now turn to the particular case, when the ensemble of
droplets at zero time
represents a two-scale Cantor dust. Recall that
the initiator of this fractal is an $E$-dimensional
cube of unit side length. The generator consists
of $n_{1}$ cubes of side $l_{1}$ and $n_{2}$
cubes of side $l_{2}$ where $l_{2}>l_{1}$.
In each step of the fractal construction every full cube
is replaced by the properly rescaled generator.
After the last step of the construction, which is the
$m$-th step, all the cubes are replaced by spherical droplets
with the same size as the cubes.

Now assume that this two-scale Cantor dust undergoes the
simple coarsening dynamics described earlier. For convenience,
we will compute
the time-dependent $d$-measure of a two-scale Cantor dust which
consists of cubes (the ones which were
replaced by the spheres after the $m$-th generation
of the construction). The only difference in the computed
$d$-measure will be in a $d$-dependent prefactor.
Since this prefactor is independent of
$k$ and $m$, it will not affect the dynamical behavior
of the $d$-measure and the Hausdorff dimension.

The $d$-measure of the $m$-th generation
of a two-scale Cantor dust can be written as \cite {Peleg}
\begin{eqnarray}
M_{d}=\int_{0}^{m} n_{m}(k)R_{m}^{d}(k)\,dk = \nonumber \\
\left(\frac{m}{2\pi k(m-k)}\right)^{1/2}
\int_{0}^{m} \exp \left[g(k)\right]\,dk
\,,
\label{12}
\end{eqnarray}
where
\begin{equation}
g(k)=-k\ln \left(\frac{k}{mn_{2}l_{2}^{d}}\right)
-(m-k)\ln \left(\frac{m-k}{mn_{1}l_{1}^{d}}\right)
\,,
\label{13}
\end{equation}
and $R_{m}(k)=l_{1}^{m-k}l_{2}^{k}$ is the size of the cubes
in the subset $S_{m}(k)$.
The function $\exp [g(k)]$ has a (sharp) maximum at
\begin{equation}
\tilde {k_{0}}(d)=\frac{n_{2}l_{2}^{d}m}
{n_{1}l_{1}^{d}+n_{2}l_{2}^{d}}
\,.
\label{14}
\end{equation}
For $d=D_{0}$ one can show that
$\tilde{k_{0}}(D_{0})=k(\alpha_{0})\equiv k_{0}$.
At time $\tau=k_{min}$ the $d$-measure of the object is
\begin{equation}
M_{d}(\tau)=\int_{k_{min}(\tau)}^{m} n_{m}(k)R_{m}^{d}(k)\,dk
\,.
\label{15}
\end{equation}
As long as $k_{min}(\tau) \ll \tilde{k_{0}}(d)$,
one can apply the saddle point argument
used in Ref. \cite {Peleg} and conclude that
\begin{equation}
M_{d}(\tau \ll \tilde{k_{0}}(d))\simeq M_{d}(\tau=0)=
(n_{1}l_{1}^{d}+n_{2}l_{2}^{d})^{m}
\,.
\label{16}
\end{equation}
This implies that during the early stages of the dynamics
the $d$-measure remains, with an exponential accuracy, constant.
Correspondingly,
the Hausdorff dimension which is computed by solving the same equation
\begin{equation}
n_{1}l_{1}^{d}+n_{2}l_{2}^{d}=1
\label{16a}
\end{equation}
for $d$, does not change with time. This
is in agreement with Eq. (\ref{11}) obtained in
the general case.

On the other hand, when $\tilde{k_{0}}(d) \ll \tau=k_{min} \ll m$,
the behavior of $M_{d}(\tau)$ is
quite different. Since for $k > \tilde{k_{0}}(d)$ $g(k)$
is a decreasing function of $k$, the main contribution
to the integral in Eq. (\ref{15}) comes from a close
neighborhood of $k=k_{min}(\tau)$. Therefore, in Eq. (\ref{15})
we can expand $g(k)$ around $k=k_{min}(t)$ to the first order
and get
\begin{eqnarray}
M_{d} \simeq
\frac{n_{m}(k_{min})R_{m}^{d}(k_{min})}{\mid g'(k_{min})\mid}= \nonumber \\
\frac{h(\xi_{min},d)\,[y(\xi_{min},d)]^{m}}{{m}^{1/2}}
\,,
\label{17}
\end{eqnarray}
where $\xi_{min}=k_{min}/m$,
\begin{equation}
y(\xi_{min},d)=\left(\frac{1-\xi_{min}}{n_{1}l_{1}^d}\right)
^{\xi_{min}-1}
\left(\frac{\xi_{min}}{n_{2}l_{2}^d}\right)
^{-\xi_{min}}
\,,
\label{18}
\end{equation}
and
\begin{equation}
h^{-1}(\xi_{min},d)=[2\pi \xi_{min}(1-\xi_{min})]^{1/2}
\mid \ln \left[\frac{(1-\xi_{min})n_{2}l_{2}^{d}}
{\xi_{min}n_{1}l_{1}^{d}}\right] \mid
\,.
\label{19}
\end{equation}
The Hausdorff dimension of the subset labeled by $\xi_{min}$
is given by
\begin{equation}
f(\alpha(\xi_{min}))= \frac
{\xi_{min}\ln(\frac{\xi_{min}}{n_{2}})+
(1-\xi_{min})\ln(\frac{1-\xi_{min}}{n_{1}})}
{(1-\xi_{min})\ln l_{1}+\xi_{min}\ln l_{2}}
\,.
\label{20}
\end{equation}
It follows that
\begin{equation}
R_{m}(k_{min})^{-f(\alpha(k_{min}))}=
\frac{n_{m}(k_{min})}{\left(\frac{m}
{2\pi k_{min}(m-k_{min})}\right)^{1/2}}
\,.
\label{21}
\end{equation}
Hence, we obtain the following expression for $M_{d}$
in the limit of $\tilde{k_0}(d) \ll \tau \ll m$:
\begin{equation}
M_{d} \simeq
\left[\frac{h(\xi_{min},d)}{m^{1/2}} \right]
R_{m}(k_{min})^{d-f(\alpha(k_{min}))}
\,.
\label{22}
\end{equation}
We see that, up to logarithmic corrections
resulting from the factor $h(\xi_{min},d)$,
the $d$-measure obeys a power law of $R_m (k_{min})$ with
a time-dependent exponent.

Eqs. (\ref{17})-(\ref{19}) allow one to calculate the
Hausdorff dimension of the ensemble of droplets in the
limit of $\tilde{k_0}(d) \ll \tau \ll m$. Taking
the logarithm
of both sides of Eq. (\ref{17}) and dividing by $m$, we get
\begin{equation}
\frac{\ln M_{d}}{m} \simeq
\frac{1}{m}\ln \left[\frac{h(\xi_{min},d)}{m^{1/2}}\right]
+\ln [y(\xi_{min},d)]
\,.
\label{23}
\end{equation}
As $m \gg 1$, the first term on the right hand side of Eq.
(\ref{23}) can be neglected. Therefore, the Hausdorff
dimension is determined by solving the equation
\begin{equation}
y\,(\xi_{min},d)=1
\,
\label{24}
\end{equation}
for $d$. The solution is just the Hausdorff
dimension of the subset $\xi_{min}$ given by Eq. (\ref{20}).
Therefore,
$D(\tau)=f(\alpha(k_{min}))$  for $k_{0} \ll \tau \ll m$,
in agreement with the general result (\ref{8}).

We now show that the assumptions of a discrete time and of the
independent shrinking of the droplets can be relaxed. It is sufficient to
 assume only that the dynamics of
each droplet are determined by its radius (and possibly by a
time-dependent ``critical radius", characterizing some mean-field
interaction between
droplets). We should also assume that the droplets
do not merge or break up. Under these assumptions the number of droplets
in each subset is constant (until the droplets 
disappear) and all the droplets belonging to the same subset have
the
same (time-dependent) radius. In addition, we forbid nucleation,
 which is a standard assumption for a coarsening stage \cite{Gunton}.

Let us denote the radii of the droplets belonging to the $k$-th
subset at time $t$ by $R_{m}(k,t)$.
The $d$-measure of the $k$-th subset at time $t$ is given by :
\begin{equation}
M_{d}(m,k,t)=n_{m}(k)R_{m}^{d}(k,t)
\,.
\label{26}
\end{equation}
This can be rewritten as :
\begin{equation}
M_{d}(m,k,t)=M_{d}(m,k,0)\left[
\frac {R_{m}(k,t)}{R_{m}(k,0)}
\right]^{d}
\,,
\label{27}
\end{equation}
where $R_{m}(k,0)$ and $M_{d}(m,k,0)$ are the initial values of the
radii and $d$-measure.
Since the initial condition is a geometrical multifractal,
$M_{d}(m,k,0)$ can be expressed in the following manner :
\begin{equation}
M_{d}(m,k,0)=\left[Y(\frac{k}{m},d,\{P_{i}\})
\right]^{m}
\,,
\label{28}
\end{equation}
where the function $Y$ and the parameters $\{P_{i}\}$ characterize 
the initial fractal
condition considered. (In our example
of the two-scale Cantor dust the role of the function $Y$ was 
played by $y$, while the set of parameters $\{P_{i}\}$
included $n_1, n_2, l_1$ and $l_2$.) 
Substituting (\ref{28}) into (\ref{27}), taking
the logarithm of both sides, and dividing by $m$ we get
\begin{equation}
\frac {\ln M_{d}(m,k,t)}{m}=
\ln \left[Y(\frac{k}{m},d,\{P_{i}\})\right] +
\frac{d}{m}\ln\left[
\frac {R_{m}(k,t)}{R_{m}(k,0)}
\right]
\,.
\label{29}
\end{equation}
For typical coarsening mechanisms $R_{m}(k,t)$ grows
with time slower than exponentially.
For example, this is true for non-conserved dynamics (model A)
and for the
Lifshitz-Slyozov theory of conserved dynamics (model B) \cite{Gunton}.
Therefore, when $R_{m}(k,t)>R_{m}(k,0)$ the second term on the right
side of Eq. (\ref{29}) is negligible at $m \gg 1$. Similarly,
it is negligible when
$R_{m}(k,t)<R_{m}(k,0)$ as long as $R_{m}(k,t)$
is not exponentially smaller than $R_{m}(k,0)$. Eq. (\ref{29}) becomes
inconvenient in the case of shrinking droplets
at the moment of their disappearance.
Eq. (\ref{27}) shows, however, that the $d$-measure of 
such droplets vanishes.
Hence, the $d$-measure of the
$k$-th subset does not change during the coarsening dynamics
until the droplets belonging to this subset disappear.
Consequently, the Hausdorff dimension of this subset
does not change until its disappearance. We have therefore shown
that the results of our simple discrete time coarsening model
apply to a wide range of coarsening mechanisms.
It should be noticed that for a system with weak multifractal
properties our model predicts that the fractal dimension remains
approximately constant at all times. Therefore, this
model provides a simple
explanation to the observation that the fractal dimension does not change
in a wide range of coarsening processes
\cite{Jullien1,Irisawa1,Conti,Irisawa2,Jullien2,Streitenberger,Crooks}.

In summary, we have considered a simple  model of coarsening
disconnected droplets
forming  a geometrical
multifractal.
We have shown that at early times the Hausdorff dimension of the system
does
not change, whereas at late times its dynamics follow the
$f(\alpha)$ curve of the initial multifractal distribution.
These results are insensitive to
the particular coarsening mechanism. We hope that they
will motivate experimental investigation of multifractal
aspects of fractal coarsening.

This work was supported in part by a grant from Israel Science
Foundation, administered by the Israel Academy of Sciences and
Humanities.

\end{document}